\documentclass[12pt]{article}
\usepackage[margin=1.0in]{geometry}
\usepackage[utf8]{inputenc}
\usepackage{amsthm,amsmath,physics,mathtools,amssymb}

\usepackage[
backend=biber,
style=phys,
biblabel=brackets,
sorting=none
]{biblatex}
\usepackage{charter}

\usepackage{CJKutf8}

\usepackage{enumitem}

\usepackage{hyperref}
\hypersetup{
    colorlinks=true,
    linkcolor=blue,
    filecolor=magenta,      
    urlcolor=cyan,
}
\usepackage[capitalise]{cleveref}

\theoremstyle{definition}
\newtheorem{theorem}{Theorem} 
\newtheorem{proposition}[theorem]{Proposition} 

\usepackage{graphicx}
\graphicspath{ {images/} }

\usepackage{tensor}

\DeclarePairedDelimiterX{\inp}[2]{\langle}{\rangle}{#1, #2}

\usepackage{xparse}

\NewDocumentCommand\LH{mo}{%
  \IfNoValueTF{#2}
   {\mathcal{L}(\mathcal{H}^{#1})}
   {\mathcal{L}(\mathcal{H}^{#1},\mathcal{H}^{#2})}%
}

\newcommand\id{\leavevmode\hbox{\small1\kern-3.3pt\normalsize1}}

\allowdisplaybreaks

\usepackage[T1]{fontenc}

\usepackage{bbold}

\usepackage{authblk}

\usepackage[title]{appendix}

\usepackage{upgreek}

\usepackage{mdframed,lipsum}
\mdfdefinestyle{MyFrame}{%
    linecolor=black,
    outerlinewidth=2pt,
    innertopmargin=4pt,
    innerbottommargin=4pt,
    innerrightmargin=4pt,
    innerleftmargin=4pt,
        leftmargin = 4pt,
        rightmargin = 4pt
        }

\title{Experience in quantum physics: toward a theory of everything}

\author{Ding Jia (贾丁)\thanks{djia@perimeterinstitute.ca}}
\affil{Perimeter Institute for Theoretical Physics, Waterloo, Ontario, N2L 2Y5, Canada}
\affil{Department of Physics and Astronomy, University of Waterloo, Waterloo, Ontario, N2L 3G1, Canada}
\date{}

\begin{document}

\begin{CJK*}{UTF8}{gbsn}
\maketitle
\end{CJK*}

\begin{abstract}
A theory of everything should not only tell us the laws for matter, gravity, and possibly boundary condition for the universe. In addition, it should specify the relation between theory and experience. Here I argue for a minimal prescription in extracting empirical predictions from path integrals by showing that alternative prescriptions are unjustifiable. In this minimal prescription, the relative probability for one experience is obtained by summing over all configurations compatible with that experience, without any further restriction associated with other experiences of the same or other experiential beings. An application to Wigner's friend settings shows that quantum theory admits objective predictions for subjective experiences. Still, quantum theory differs from classical theory in offering individualized as opposed to collective accounts of experiences. This consideration of experience in fundamental theories issues several challenges to popular quantum interpretations, and points to the outstanding need for a theory of experience in understanding physical theories of everything.
\end{abstract}

\tableofcontents

\section{Introduction}\label{sec:i}

A physical theory of everything is supposed to tell us \cite{Page2003QuantumCosmology}:
\begin{enumerate}
    \item The dynamical laws for matter and gravity.
    \item The laws for the boundary condition of the universe, if there are such laws.
    \item The relation between the theory and experience.
\end{enumerate}
Task 1 is the focus of particle physics and quantum gravity. Task 2 is a main topic of quantum cosmology. Task 3 is, well, a big embarrassment of quantum physics.

\begin{figure}
    \centering
    \includegraphics[width=0.9\textwidth]{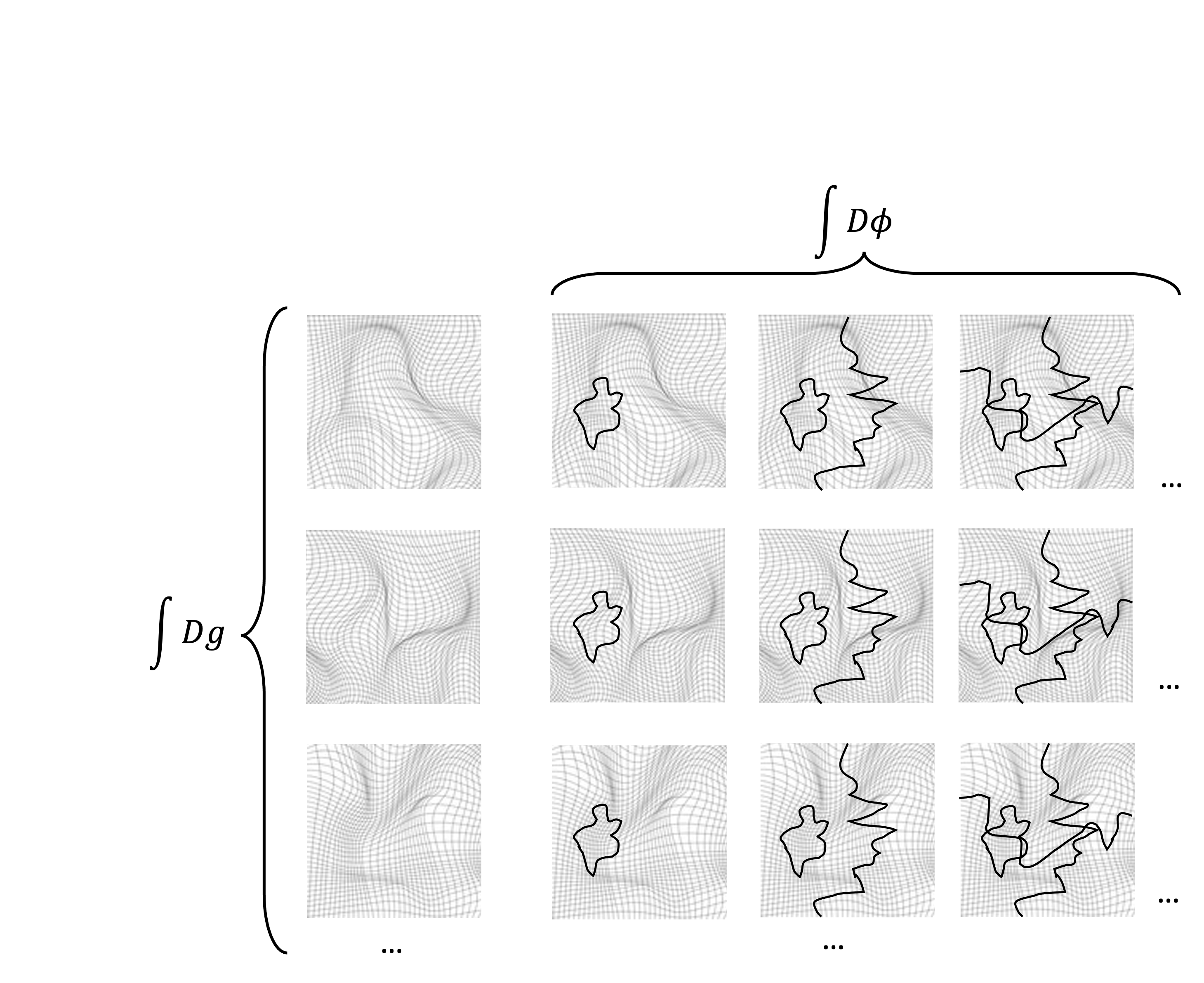}
    \caption{The path integral sum over everything $\int Dq=\int Dg \int D\phi$ contains gravity part $\int Dg$ and a matter part $\int D\phi$ (for simplicity, illustrated in the figure by particle configurations, instead of more realistic field or particle-string \cite{Jia2022WhatModel} configurations for the Standard Model).}
    \label{fig:pi_sm}
\end{figure}

Task 1 is fulfilled most straightforwardly by starting with our best theory for matter, the Standard Model, and extending its path integral to sum also over gravity (\Cref{fig:pi_sm}). As a result, we get partition functions of the form
\begin{align}
    Z=\int Dq ~ e^{\frac{i}{\hbar} S[q]},
\end{align}
where $q$ contains both matter and gravity variables and $S[q]$ is the action. The matter part may be extended, e.g. to incorporate dark matter. The gravity part has more than one possible realization. See \cite{oriti2009approaches} for a survey of some popular approaches.

Task 2 invokes double path integrals of the form (see equation (4.8) of \cite{Hartle1995SpacetimeSpacetime})
\begin{align}\label{eq:pib}
    D[\rho]=\int Dq'  \int Dq ~ 
e^{\frac{i}{\hbar}\left(S[q] -S[q']\right)}\rho(q_b, q_b'),
\end{align}
where the path integral is doubled to account for the possible mixedness of the boundary condition $\rho$, which as an analog of the density operator takes as inputs the boundary configurations $q_b$ and $q_b'$ in the double copies. In case no law exists to fix the boundary condition, $\rho$ is treated as an unknown parameter to be inferred empirically.

Task 3 is the focus of this work. This task is a big embarrassment, because we do not know exactly how the theory relates to experience quantum physics. From textbooks such as \cite{Feynman1965QuantumIntegrals} we gather that in the path integral formalism, probabilities should be obtained through selecting a subset of configurations to integrate. However, there is no instruction on how to select in a theory of everything where all matter and gravity variables are subject to path integration. In particular, to predict Alice's experience at one moment, do we just select according to that experience alone? Do we select on Alice's past and/or future experiences as well? Do we also select on Bob's and others' experiences, etc.? 

These questions are important for cosmology, for instance in understanding how empirical probabilities are obtained in cosmology \cite{Castagnino2017InterpretationsCosmology} and in resolving the issue of Boltzmann brain \cite{Carroll2017WhyBad}. The questions are important for foundations of quantum field theory, for instance in understanding signalling constraints of measurements in QFT \cite{Sorkin1993ImpossibleFields, Borsten2021ImpossibleRevisited, Jubb2022CausalTheory}. The questions are important for foundations of quantum physics, for instance in understanding Wigner's friend settings \cite{Wigner1961RemarksQuestion}. 

In this paper, I argue for a ``minimal prescription'' for extracting empirical predictions in theories of everything. The prescription is ``minimal'', because in predicting the probability for an experience, the path integral selects only for that experience. Alternative prescriptions which select for other experiences are systematically discussed, and are shown to be unjustifiable. 

To illustrate the minimal prescription, I apply it to Wigner's friend \cite{Wigner1961RemarksQuestion} type thought experiments and show how touching base with fundamental theories fixes a unique and general prescription that accounts for the experiences of Wigner, Friend, and any other experiential beings. I explain why no commitment to any quantum interpretation is needed to arrive at the prescription, and why the recent Wigner's friend no-go theorems \cite{Brukner2017OnProblem, Brukner2018AFacts, Frauchiger2018QuantumItself, Pusey2016Is20, Bong2020AParadox, Zukowski2021PhysicsResults} cannot help in arriving at the prescription.

Quantum theory with the minimal prescription gives an objective account for experiences, since everyone applies the same formula to account for anyone's experiences. The account is individualized, in the sense that a different formula is needed for each different experience. This is in contrast to classical theory, where a collective account of multiple experiences with the same formula is possible. In the world picture that emerges from the quantum theory of everything, matter and gravity path integral configurations coexist in superposition, whereas experience induces selection. Quantum states have no fundamental status, since the theory contains only boundary condition and experience selections, but not states that evolve and get updated in time. To the extent that experience is part of ``reality'', both boundary condition and experience selection describe ``reality''. 

Based on these observations, I discuss some shortcomings of pilot-wave theories, collapse models, as well as Everttian, decoherent histories, relational, QBism, and neo-Copenhagen interpretations. Judged in the context of theory of everything, these approaches are in danger of being wrong, redundant, vague, and/or even superfluous. It remains to be seen if they offer any help in completing the remaining parts of three tasks for a theory of everything mentioned at the beginning.

As a note of scope, the discussion of theory of everything in this work is based in the path integral formalism.\footnote{In my view, the most promising candidate theories of quantum gravity, e.g., Lorentzian simplicial quantum gravity \cite{Jia2022ComplexProspects}, locally causal dynamical triangulation \cite{Jordan2013CausalFoliation} etc., come from the path integral formalism.} Readers interested in canonical/algebraic formulations may find Don Page's ``Sensible Quantum Mechanics'' \cite{PageSensibleProbabilistic, Page1995SensibleMind, Page2003MindlessConsciousness} relevant in accounting for experiences. Page's formalism is closely related to this paper in attributing quantum probabilities to individual experiences. However, it gives an essentially different prescription for making predictions in cosmology, where ambiguities associated with Boltzmann brain constitute a genuine problem \cite{PageConsciousnessQuantum, PageBornsUniverse}. In contrast, the treatment presented here holds Boltzmann brain problem as a fake problem, as I will elaborate on elsewhere.

An outline of the paper can be found in the table of contents at the beginning.

\section{Experience in quantum physics}\label{sec:eqp}

\subsection{From particle to everything}

\begin{figure}
    \centering
    \includegraphics[width=0.75\textwidth]{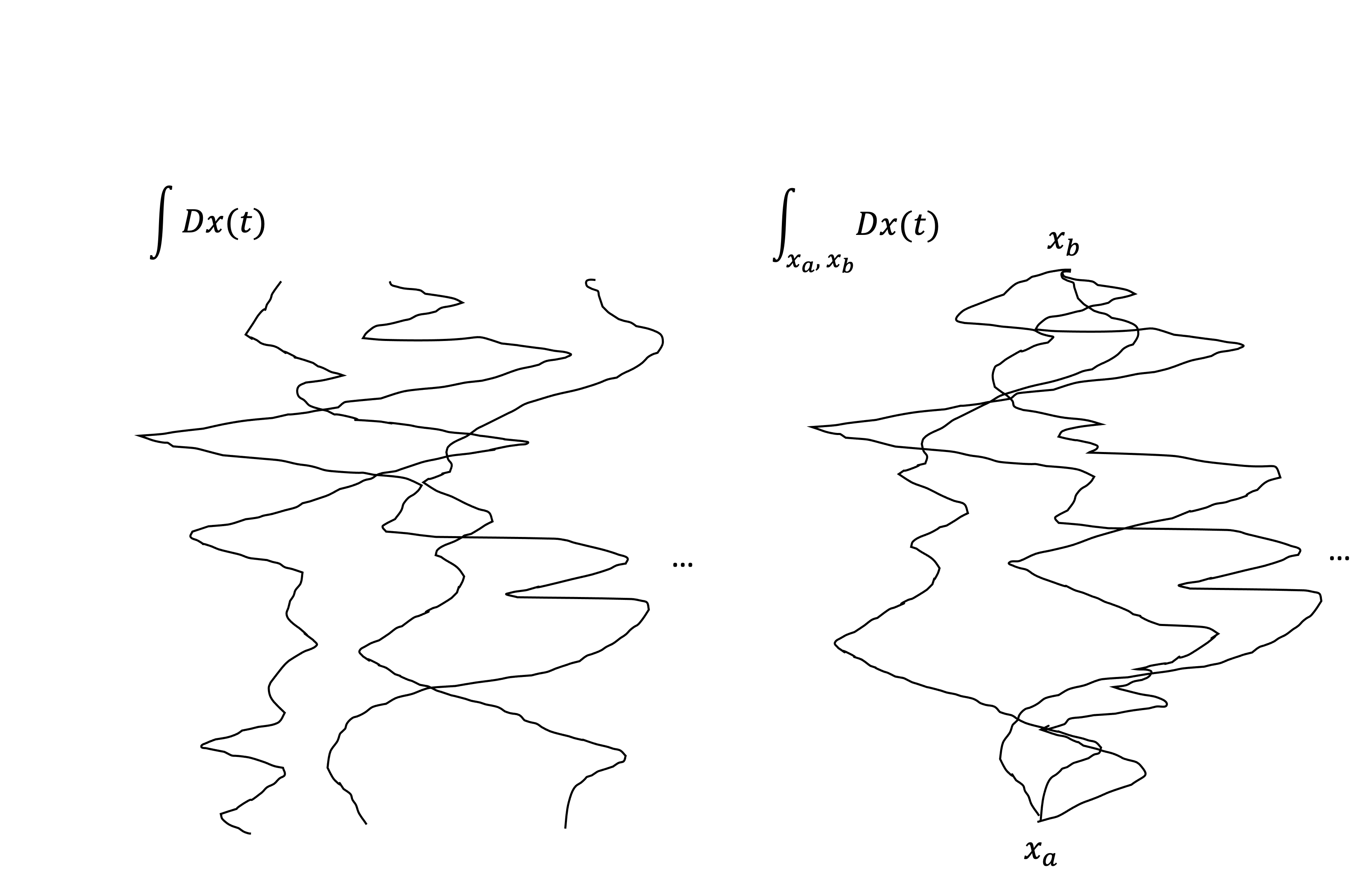}
    \caption{From the set of all particle paths (left), select those paths compatible with observational locations (right) to derive observational probabilities.}
    \label{fig:selection_particle}
\end{figure} 

How should we extract empirical predictions from a path integral? In chapters 1 and 2 of \cite{Feynman1965QuantumIntegrals}, Feynman and Hibbs offer a textbook treatment for a non-relativistic particle. To compute the probability for emitting and detecting a particle at a certain locations, one sums the amplitude over all paths originating and ending at these locations, and square it to obtain the probability (\Cref{fig:selection_particle}). In other words, we select from all particle paths those compatible with the observational locations to obtain the probability for the observation. 


\begin{figure}
    \centering
    \includegraphics[width=0.95\textwidth]{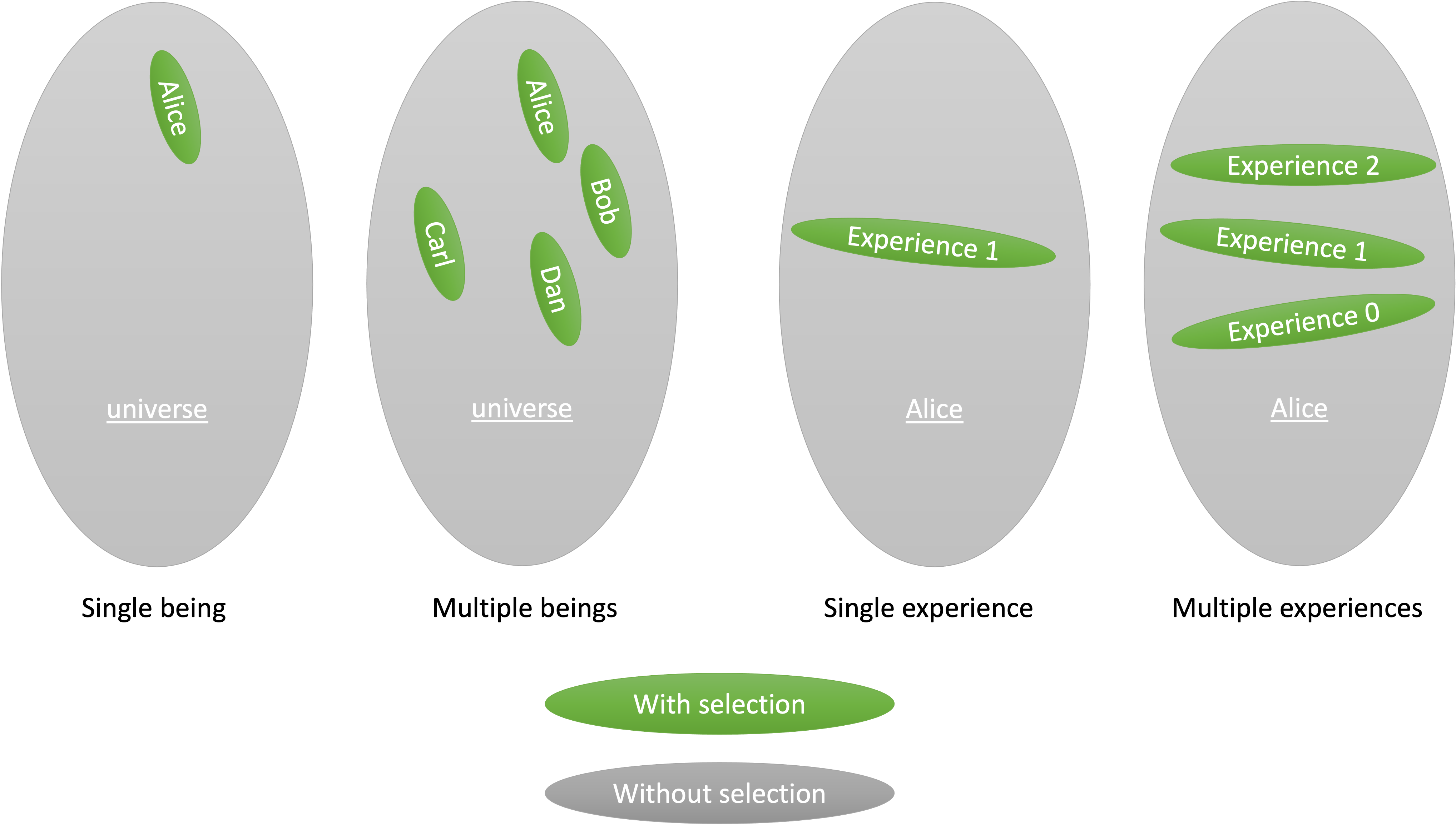}
    \caption{Different possibilities of selection for experiences.}
    \label{fig:selection_scenarios}
\end{figure} 

In accounting for experiences in a theory of everything of the form \eqref{eq:pib}, it is natural to also select the physical configurations compatible with the experience under consideration to derive its probability. However, as soon as we attempt to formalize this idea, questions arise (\Cref{fig:selection_scenarios}). There are many experiential beings in the world, each going through many experiences. When extracting the probability for one experience of one being, should we select according to the experiences of all experiential beings, or just that one being? For that particular being, should we select according to multiple experiences, or just that one experience under consideration?

The answer is hard to find in previous works, even those that pay special attention to foundational questions of path integrals. The decoherent histories understanding of matter-gravity path integrals \cite{Hartle1995SpacetimeSpacetime} suffers ambiguities in history selection, which makes it unclear how to extract empirical predictions \cite{Dowker1996OnMechanics}. The general boundary formalism \cite{Oeckl2019APhysics} does not clearly fix a prescription on how many boundary conditions to impose, even when supplied with a relational interpretation \cite{Rovelli2022TheInterpretation} (In a universe containing many experiences and experiential beings, should we apply boundary condition to one experience? Some experiences? All experiences?). No other work I know of addresses the questions either. To proceed, we must find our own way.

\subsection{General formula for experience}\label{sec:gfe}

I argue below that to extract the relative probability for some experience, we should apply selection for just that one experience. Consider the empirical probabilistic predictions in the form of conditional probabilities
\begin{align}\label{eq:epp}
    p(e_i|c).
\end{align}
Here $c$ stands for a physical condition that enables a set of possible experiences $\{e_i\}_i$. For example, $c$ could label the momentary (which may extend in time) physical configuration of a human being, incorporating all matters relevant for what he/she experiences next. In principle, this should fix the set $\{e_i\}_i$ of all possible next experiences.

In the \textbf{minimal prescription}, we select for a single experience for a single experiential being in the path integral. This corresponds to following the scenario of the first and third pictures of \Cref{fig:selection_scenarios}. The probability for $e_i$ to actualize under condition $c$ is
\begin{align}\label{eq:mf}
    p(e_i|c)=N \int_{e_i,c} Dq'  \int_{e_i,c} Dq ~ 
e^{\frac{i}{\hbar}\left(S[q] -S[q']\right)}\rho(q_b, q_b').
\end{align}
The subscripts $e_i,c$ indicate that in both branches of the double path integral, we select physical configurations compatible with the condition $c$ and the experience $e_i$. The normalization constant $N$ is fixed by requiring $\sum_i p(e_i|c)=1$. 

There are several open questions on how to implement the selection exactly. For instance, it is unclear what physical conditions enable experiences, what physical configurations correspond to experiences, and what the list of all possible experiences exactly is under a given condition \cite{Chalmers1995FacingConsciousness, Dennett2018FacingConsciousness}. It is unclear if human experience supervenes on just the brain physical configurations, or if some part of the bodily configurations are also relevant \cite{Shapiro2021EmbodiedCognition}. It is unclear how much the physics of experience varies from species to species \cite{Godfrey-Smith2016OtherConsciousness}. It is unclear at the mathematical level if the selections should always be sharp (a configuration is either included or excluded in the path integral), or can be unsharp (multiply the configurations by a weighting function taking non-binary values). ...

These are important scientific questions needing multidisciplinary inputs. On the other hand, just the form of the formula \eqref{eq:mf} without the details already has rich implications in physics. In the following, I will focus on extracting some of these implications.

\subsection{Why select just one experience?}\label{sec:woe}

Before moving on to the implications, I should explain why we should select according to just one experience in the probability formula \eqref{eq:mf}. To start with, note that:
\begin{mdframed}[style=MyFrame,nobreak=true,align=center,userdefinedwidth=30em]
\centering Joint experience is not experience
\end{mdframed}
Consider the combination of some experience of Alice with some experience of Bob, or with some other experience of Alice herself. The result is not an experience experienced by any being. Therefore to make probabilistic predictions for one experience, it is reasonable to consider selecting on that experience and that experience only.


\begin{figure}
    \centering
    \includegraphics[width=1.0\textwidth]{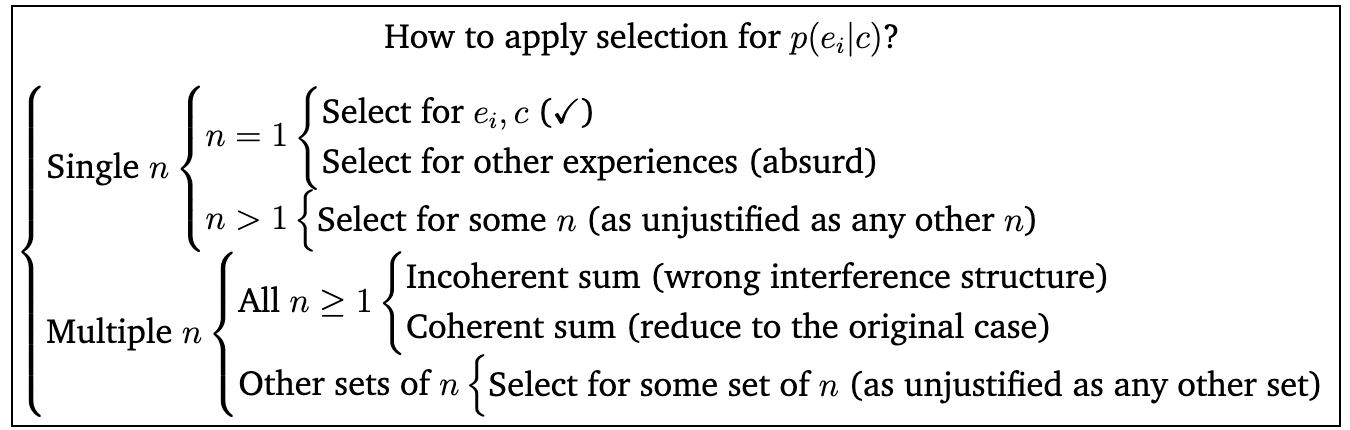}
    \caption{Structure of argument for picking the selection prescription for $p(e_i|c)$. Here $n$ is the number of experiences to be selected for.}
    \label{fig:selection_prescriptions}
\end{figure} 

In contrast, alternative selection scenarios can be shown to be quite unreasonable (\Cref{fig:selection_prescriptions}). If we were to select just for $n=1$ experience, it is certainly reasonable to select according to one under consideration as in \eqref{eq:mf}, instead of a different experience. If we were to select for $n>1$ experiences, there is no $n$ that is preferred for any reason. Therefore any particular choice of $n$ lacks a good justification. 

One might ask, why do we not select according to the number of experiences, or the number of experiential beings, that exist in the universe? The problem is that there is no fixed number of experiences or experiential beings in the path integral over everything \eqref{eq:pib}. In any quantum region \cite{Oeckl2019APhysics}, the path integral sums over gravity and matter configurations with zero, one, two ... experiences, so it is meaningless to talk about a definite number of experiences or experiential beings that exist in the universe.\footnote{The boundary condition cannot help in selecting some particular $n$ value or some set of $n$ values of experiences that exist in the path integral configurations, because it only refers to the boundary part of the path integral configurations, but not the interior part where paths with different numbers of experiences are summed over.} 

The above analysis indicates that $n=1$ is the only reasonable option, if a single number of experiences is to be selected. An alternative is to select for a range of $n$ values. In this case, the first option is to sum over all $n\ge 1$. Here  $n=0$ gets excluded since we want to make sure that the experience under consideration is selected. The sum could be performed either coherently or incoherently. A coherent sum over amplitudes for all numbers $n$, keeping the selection for the experience under consideration fixed in all the terms of the sum, actually gives back \eqref{eq:mf}. On the other hand, an incoherent sum over probabilities for all numbers $n$ 
is not a reasonable choice. This is because all quantum regions contain experiential configurations, so the result will exhibits a drastically different interference structure whose predictions will differ from empirical evidence.
The second option is to sum over a different set of multiple $n$ values. The problem, like in the case of selecting for a single $n>1$, is that within quantum theory there is no particular set of $n$ that is preferred for any reason. Therefore any choice lacks a good justification. 

If one is willing to go beyond quantum theory, there could be mechanisms that fix one or a set of $n$ values. For instance, the modified dynamics of consciousness collapse models \cite{ChalmersConsciousnessFunction, Kremnizer2015IntegratedCollapse, Okon2020AModel, OkonTheChallenges, Kent2021CollapseConsciousness} may fix some $n$ value(s) through its stochastic reduction. In this paper, I focus on accounting for experience within quantum theory, so will not discuss such possibilities further (however, see \Cref{sec:qi} for some critical remarks on collapse models). 

\subsection{A toy model}\label{sec:tm}

\begin{figure}
    \centering
    \includegraphics[width=0.75\textwidth]{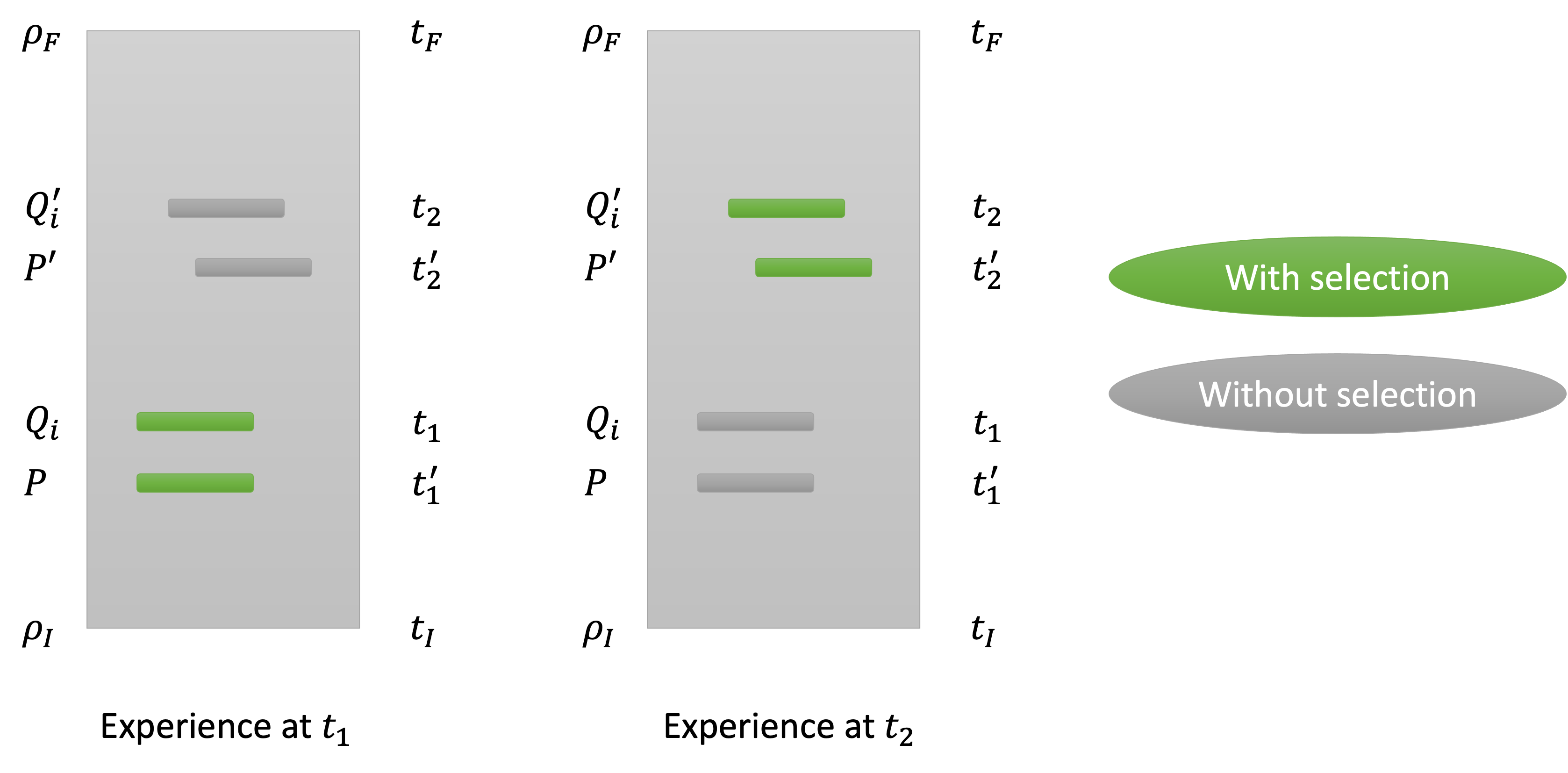}
    \caption{Different projections are activated for different experiences.}
    \label{fig:selection_tm}
\end{figure} 

Formula \eqref{eq:mf} is rather general and abstract. It helps to illustrate it in a concrete setting with three simplifying assumptions. (1) Consider a simplified setting where spacetime is reduced to classical. For instance, consider  a saddle approximation to the gravitational path integral, where one classical solution to the gravitational boundary value problem dominates the path integral. (2) Moreover, assume that the matter degrees of freedom obey a unitary time evolution for some time foliation of the classical spacetime. 
(3) Assume further that the spacetime is sufficiently non-degenerate, so that we can locate experiences sequentially in time by selecting on the gravitational configurations. 

Now consider a set of experiences $\{e_i\}_i$ possible under the condition $c$. Assume the latter to be implemented by the projector $P$ at time $t_1'$, and the former to be implemented by the projectors $Q_i$ at time $t_1>t_1'$ (\Cref{fig:selection_tm}). Then
\begin{align}\label{eq:tmpf}
    p(e_i|c)=&N \Tr[\rho_F U(t_F,t_1) Q_{i} U(t_1,t_1') P U(t_1',t_I) \rho_I U^\dagger(t_1',t_I) P U^\dagger(t_1,t_1') Q_{i}  U^\dagger(t_F,t_1)],
\end{align}
where $N$ is fixed by setting $\sum_i p(e_i|c)=1$, and as a further simplifying assumption, the boundary condition factorizes into $\rho_I$ and $\rho_F$ on the initial and final boundary of the universe. The unitaries $U(t_b,t_a)$ between times $t_a$ and $t_b$ is to be derived from the path integral propagators. In the Heisenberg picture where the operators
\begin{align}
    O(t)= U^\dagger(t,t_0) O U(t,t_0)
\end{align}
are time-dependent ($t_0$ is some reference time), the formula becomes
\begin{align}\label{eq:tmpfh}
    p(e_i|c)=&N\Tr[\rho_F(t_F) Q_{i}(t_1) P(t_1') \rho_I(t_I) P(t_1') Q_{i}(t_1)],
\end{align}
where the initial and final boundary conditions lie at times $t_I$ and $t_F$.

Suppose we want to model the experience of the same being with memory, at a later time $t_2>t_1$ (\Cref{fig:selection_tm}). Then we only impose selection for that experience to obtain 
\begin{align}\label{eq:tmpfh2}
    p(e_i'|c')=&N'\Tr[\rho_F(t_F) Q'_{i}(t_2) P'(t_2') \rho_I(t_I) P'(t_2') Q'_{i}(t_2)],
\end{align}
where $c'$ is a different condition with projector $P'$ at $t_2'\le t_2$, and $\{e'_i\}_i$ is a different set of experiences with projectors $Q'$ at $t_2$. 

The memory is encoded in $P'$. For instance, when the initial boundary condition is pure, i.e., when $\rho_I(t_I)=\ketbra{\psi}$, the memory of experience $e_j$ at $t_1$ and condition $c$ at $t_1'$ can be incorporated by setting $P'(t_2')=\ketbra{\psi_j}$ in the Heisenberg picture. Here
\begin{align}\label{eq:emp}
    \ket{\psi_j}=P''(t_2') Q_{j}(t_1) P(t_1')\ket{\psi},
\end{align}
where $P''(t_2')$ implements further conditioning not already implemented by $Q_{j}(t_1) P(t_1')$. Plugging it in \eqref{eq:tmpfh2} yields
\begin{align}\label{eq:pem}
    p(e_i'|c')\propto \Tr[\rho_F(t_F) Q'_{i}(t_2) P''(t_2') Q_{j}(t_1) P(t_1') \rho_I(t_I) P(t_1') Q_{j}(t_1) P''(t_2') Q'_{i}(t_2)].
\end{align}
Even when selections are made only at $t_2'$ and $t_2$, the influence of previous events are reflected in the current condition $c'$  because of the ``memories'' encoded in $P'(t_2')$.

This toy model only gives a rough account for the experiences of experiential beings with memory. In a more realistic account, the condition encoding memory, $P'$, should refer to the particular local bodily physical configurations of the experiential being, so will take a different from than that given by \eqref{eq:emp}. This unrealistic aspect should be kept in mind when using the toy model.

\section{Wigner's friend}\label{sec:wf}

\begin{figure}
    \centering
    \includegraphics[width=0.9\textwidth]{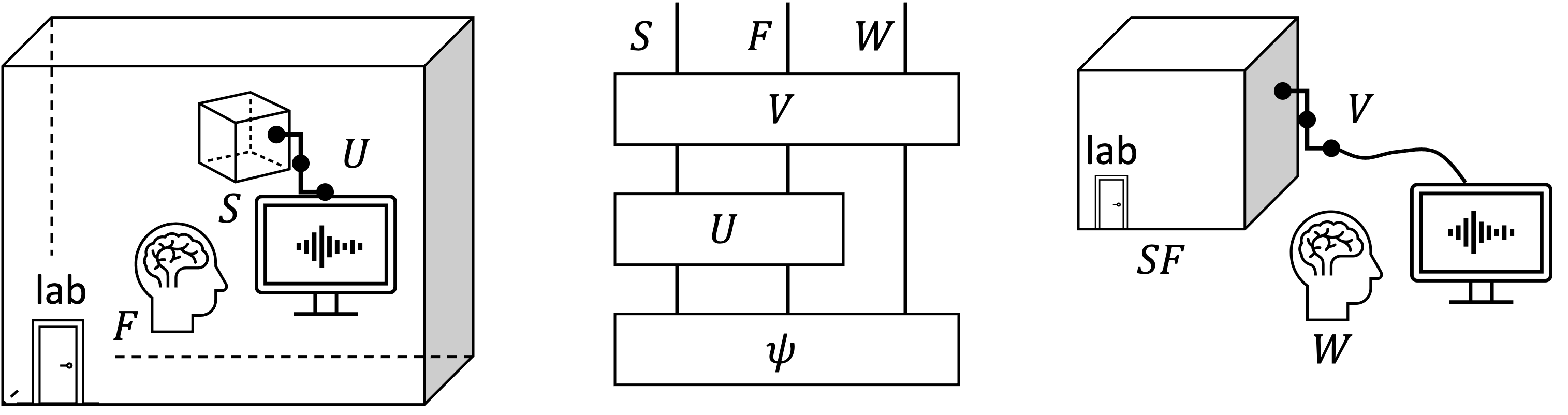}
    \caption{Wigner's friend setting. Left: Inside a lab, Friend ($F$) interacts with a system ($S$) through unitary evolution $U$. Right: Afterwards, Wigner ($W$) outside the lab interacts with system-Friend through unitary evolution $V$. Middle: circuit diagram for the interactions.}
    \label{fig:wigner_friend1}
\end{figure} 

Wigner's friend thought experiment \cite{Wigner1961RemarksQuestion} (\Cref{fig:wigner_friend1}) poses the question whether in describing the experience of one being, other beings could be in superposition of physical configurations corresponding to different experiences. This setting provides an ideal ground to illustrate the prescription of the previous section, which answers the question with a resounding yes. 

\subsection{Setting}\label{sec:s}

Consider the Wigner's friend setting of \Cref{fig:wigner_friend1} with the global unitary evolution
\begin{align}
    \ket{\psi(0)}&=\frac{1}{\sqrt{2}} (\ket{0}_S+\ket{1}_S)\ket{f}_F\ket{w}_W
    \\
    \xrightarrow{U}\ket{\psi(1)}&=\frac{1}{\sqrt{2}} (\ket{00}_{SF}+\ket{11}_{SF})\ket{w}_W
    \\
    \xrightarrow{V}\ket{\psi(2)}&=\frac{1}{\sqrt{2}} (\ket{00}_{SF}+\ket{11}_{SF})\ket{0}_W. \label{eq:psi2}
\end{align}
At time $t=0$, the system, describing for instance the spin of a particle, is initialized in the state $\frac{1}{\sqrt{2}} (\ket{0}_S+\ket{1}_S)$. Friend's and Wigner's physical configurations are initialized in some states $\ket{f}_F$ and $\ket{w}_W$. 

At time $t=1$, the global state evolves to $\ket{\psi(1)}$ under $U$, which couples Friend to the system. This constitutes the premeasurement unitary interaction for Friend to measure the system with the projections $\{\ketbra{0},\ketbra{1}\}$. 

At time $t=2$, the global state evolves to $\ket{\psi(2)}$ under $V$, which couples Wigner to system-Friend. This constitutes the premeasurement unitary interaction for Wigner  to measure system-Friend with the projections $\{\ketbra{\Phi},\id-\ketbra{\Phi}\}$, where
\begin{align}\label{eq:Phi}
    \ket{\Phi}=\frac{1}{\sqrt{2}} (\ket{00}_{SF}+\ket{11}_{SF}).
\end{align}

Without any experiential selections, the unitary evolution does not yield probabilities for any experience. The question is to decide on the projections to imply in accounting for the experiences of Wigner and Friend.

\subsection{Wigner's experiences}\label{sec:we}



Suppose in a simplified setting all assumptions leading to the toy model of \Cref{sec:tm} hold, so we can employ the model there. Let us suppose further that Wigner's experiences are characterized by projectors $Q_i$, and the condition only retains the most recent memory by applying the projection at the closest previous time (this amounts to implementing the condition by $P'$ of \eqref{eq:emp} and setting the Heisenberg picture projectors $P''$ and $P$ to identity in \eqref{eq:emp}). Suppose further that
\begin{align}
    \rho_F=I,
\end{align}
i.e., the final boundary condition is the identity in \eqref{eq:tmpf}.

Then \eqref{eq:tmpfh} yields, in the Schr{\"o}dinger picture,
\begin{align}\label{eq:pjiW1}
    p(j|i)=&N \Tr[Q_j U Q_i\ketbra{\psi(0)}Q_i U^\dagger Q_j], \quad t=1,
    \\
    \label{eq:pjiW2}
    p(j|i)=&N \Tr[Q_j V Q_i\ketbra{\psi(1)}Q_i V^\dagger Q_j], \quad t=2.
\end{align}
for experience $j$ conditioned on memory for the previous experience $i$. 
For example, when there are only two possible experiences with projectors
\begin{align}
    \{Q_0=\id_{SF}\otimes\ketbra{0}_W,Q_1=\id_{SF}\otimes\ketbra{1}_W\},
\end{align}
plugging in \eqref{eq:pjiW1} yields that at time $t=1$, $p(j|i)=\delta_{i,j}$, where the normalization factor $N$ is fixed to be $|\bra{i}\ket{w}|^2=1$. This indicates that at $t=1$ the experience must stay the same as previously.

At $t=2$, the explicit form of $V$ is needed to compute the result for \eqref{eq:pjiW2}. As an example, let $W$'s state space be spanned by $\{\ket{w}, \ket{w_\perp}\}$ with
\begin{align}
    \ket{i}=\alpha_i \ket{w}+\beta_i \ket{w_\perp}, \quad k=0,1,
\end{align}
and $F$'s state space be spanned by $\{\ket{\Phi},\ket{\Phi_1},\ket{\Phi_2},\ket{\Phi_3}\}$ with $\ket{\Phi}$ given in \eqref{eq:Phi}. Suppose
\begin{align}
    V: &\ket{\Phi w}\mapsto \ket{\Phi 0}, \ket{\Phi_k w}\mapsto \ket{\Phi_k 1},\nonumber
    \\
    &\ket{\Phi w_\perp}\mapsto \ket{\Phi 1}, \ket{\Phi_k w_\perp}\mapsto \ket{\Phi_k 0}, \quad k=1,2,3.\label{eq:Vf}
\end{align}
Then indeed $V$ is compatible with \eqref{eq:psi2}. 
Plugging in \eqref{eq:pjiW2} yields
\begin{align}
    p(j|i)=& N\norm{\bra{j} V (\ketbra{w}+\ketbra{w_\perp}) \ket{\Phi i}\braket{i}{w}}^2
    \\
    =& N\abs{\delta_{j,0} \alpha_i \alpha_i^* + \delta_{j,1} \beta_i \alpha_i^*}^2
    \\
    =& \abs{\delta_{j,0} \alpha_i + \delta_{j,1} \beta_i}^2,
\end{align}
where the normalization factor $N$ has been fixed to be $\abs{\alpha_i}^2$. This indicates that the probabilities for experience $j$ at $t=2$ are $\abs{\alpha_i}^2$ for $j=0$ and $\abs{\beta_i}^2$ for $j=1$.

\subsection{Friend's experiences}\label{sec:fe}

The formulas for Friend's experiences are entirely analogous. Suppose Friend's experiences are characterized by projectors $R_i$ (e.g., $\{R_0=\id_{SW}\otimes\ketbra{0}_F,R_1=\id_{SW}\otimes\ketbra{1}_F\}$).
Then the probabilities take the same form as Wigner's \eqref{eq:pjiW1} and \eqref{eq:pjiW2}:
\begin{align}\label{eq:pjiF1}
    p(j|i)=&N \Tr[R_j U R_i\ketbra{\psi(0)}R_i U^\dagger R_j], \quad t=1,
    \\
    \label{eq:pjiF2}
    p(j|i)=&N \Tr[R_j V R_i \ketbra{\psi(1)} R_i V^\dagger R_j], \quad t=2.
\end{align}
Once explicit forms of $U$ and $V$ like \eqref{eq:Vf} above are given, the probabilities can be computed explicitly. 

More generally, there can be alternative settings with possibly more experiential beings interacting in arbitrarily complicated ways. In all these cases, the probability formulas would take the same form as \eqref{eq:pjiW1}, \eqref{eq:pjiW2}, \eqref{eq:pjiF1} and \eqref{eq:pjiF2}. The prescription of \Cref{sec:eqp} applies in generality.

\subsection{Contrasting with previous treatments}

The present treatment of the Wigner's friend setting within quantum theory contains some differences from some previous treatments. 

In \cite{DeBrota2020RespectingFriend}, DeBrota, Fuchs, and Schack give an analysis of Wigner's friend settings from a QBism perspective. On page 1869 of \cite{DeBrota2020RespectingFriend}, one finds the statement: ``the friend [...] amounts to assigning a quantum state to herself, which violates the QBist tenet that there must be a clear separation between agent and measured system''. In contrast, the formulas in the present section do not forbid someone from assigning a state to herself. For example, when Friend makes a prediction for Wigner's experiences, like everyone else she should use the formulas of \Cref{sec:we} which assign a state to herself. Conceptually, this highlights the \textbf{objective} nature of the quantum empirical predictions, which is easily forgotten in emphasizing the subjective nature of experiences \cite{DeBrota2020RespectingFriend}.

In some treatments of Wigner's friend settings (e.g., \cite{Frauchiger2018QuantumItself, Baumann2020WignersAgent}), the humans of Wigner and Friend are replaceable by computers/automatic machines without making a difference. This is in contrast with the experiential treatment given here where experiential beings are special. In a world \textbf{without experiential beings}, there is no empirical prediction to make, and no selections to be applied to the path integral (and no projections to be applied to the state in the toy model). Therefore computers/machines without experience do not induce definite values for variables, and must be treated differently from humans with experience. 

In \cite{Baumann2020WignersAgent}, Baumann and Brukner imagines a ``textbook treatment'' of a Wigner's friend setting. In this treatment, even when predicting Wigner's observational outcomes, Friend applies projective state-updates for her own measurements. Baumann and Brukner point out that the ``textbook treatment'' differs from treatments by QBism, neo-Copenhagen, and relational interpretations within quantum theory, which leaves the impression that these \textbf{interpretations} make a genuine difference. In contrast, according to the analysis given here which touches base with \textbf{fundamental theories} with specified dynamical laws (encoded in the action for the path integral), no such interpretations are required to differ from the ``textbook treatment''. As argued in \Cref{sec:eqp}, formula \eqref{eq:mf} is the only reasonable prescription for extracting empirical predictions in the fundamental quantum theories considered here. This leads to the formulas of \Cref{sec:we}, which, in contrast to the ``textbook treatment'', apply projections only to Wigner when accounting for Wigner's experiences (by Friend, Wigner, or whoever else). 

Finally, readers who know of the recent Wigner's friend no-go theorems \cite{Brukner2017OnProblem, Brukner2018AFacts, Frauchiger2018QuantumItself, Pusey2016Is20, Bong2020AParadox, Zukowski2021PhysicsResults} may ask if these inform us anything on how to account for experiences within quantum theory. In particular, can the no-go theorems offer alternative arguments to rule out any prescription discussed in \Cref{sec:woe}? The answer is no. The structure of argument in \cite{Brukner2017OnProblem, Brukner2018AFacts, Pusey2016Is20, Bong2020AParadox, Zukowski2021PhysicsResults} is that quantum predictions violate certain Bell-type inequalities, so the assumptions leading to the inequalities cannot coexist if quantum theory is right. For this type of argument to take off, one must already know how to draw the correct quantum empirical predictions. However, knowing this would have fixed the prescription already, without considering any inequalities. Therefore the no-go theorems of \cite{Brukner2017OnProblem, Brukner2018AFacts, Pusey2016Is20, Bong2020AParadox, Zukowski2021PhysicsResults} cannot help in picking out the correct prescription for quantum empirical predictions. The structure of argument in \cite{Frauchiger2018QuantumItself} is that certain assumptions put together yield a contradiction. If any prescription discussed in \Cref{sec:woe} obeys the assumptions, that would be ruled out by \cite{Frauchiger2018QuantumItself}. However, none of the assumptions applies, since the prescriptions only assign definite values to a physical variable when there is empirical selection (without selection, all values are summed over in the path integral), whereas the assumptions of \cite{Frauchiger2018QuantumItself} assume it is meaningful to talk about definite values of physical variable in themselves without regards to experience. Therefore the no-go theorem of \cite{Frauchiger2018QuantumItself} cannot help in picking out the correct prescription for quantum empirical predictions.\footnote{One may be tempted to ask, if the no-go theorems cannot rule out any prescription for quantum empirical prediction, what do they rule out at all? The one-paragraph discussion offered here does not address this question, but only explains why the no-go theorems cannot be used to rule out prescriptions discussed in \Cref{sec:woe}. For in-depth critical discussions that do address the question, see, for instance, \cite{OkonOnOutcomes, OkonReassessingTheorems} and references therein.}

This comparison with previous treatments reveals several conceptual questions that deserve to be clarified further. In which sense is the minimal prescription of \Cref{sec:eqp} objective? In which is it not? What picture of Nature does quantum theory offer in the absence of experiential beings like us, as some ``realists'' tend to ask? In what ways do ``fundamental'' quantum theories differ from ``non-fundamental'' quantum theories? What is the role of quantum interpretations in accounting for experiences? I will address these questions in the rest of the paper.

\section{Conceptual reflections}\label{sec:ci}

\subsection{Objective vs. subjective}\label{ovs}



The empirical probabilities derived from \eqref{eq:mf} are \textbf{objective}, in the sense that everyone uses this same formula to account for anyone's experiences. For example, in the Wigner's friend setting, Wigner, Friend, and everyone else should use the formulas of \Cref{sec:we} to account for Wigner's experiences.

Certainly the experiences themselves are subjective, in the sense that Wigner does not experience Friend's experience and \textit{vice versa}. However, this does not change the objective nature of \eqref{eq:mf} in drawing empirical predictions.

An important consequence is that the quantum empirical probabilities admit an \textbf{objective propensity interpretation} (although this does not rule out other interpretations for the quantum empirical probabilities). Quantum interpretations such as QBism \cite{Fuchs2016QBism:Handbook, Fuchs2017NotwithstandingQBism, StaceyIdeasQBism} hold that quantum probabilities are Bayesian probabilities for agents' beliefs. Here it is shown that even in view of extreme settings such as Wigner's friend, a Bayesian interpretation of quantum empirical probability is not a necessity.

\subsection{Individualized vs. collective}\label{sec:ic}

There is a sense in which ``objectivity'' is weakened in \eqref{eq:mf} in comparison to classical theory. In classical theory, a \textbf{collective} account for multiple experiences is available. For instance, in a classical field theory, a field configuration throughout the spacetime is supposed to account for all experiences throughout this universe. In contrast, the quantum formula \eqref{eq:mf} offers only an \textbf{individualized} account, in the sense that a particular selection is used for each particular experience.

This individualized vs. collective dichotomy also distinguishes quantum theory based on \eqref{eq:mf} and collapse models. The distinction leads to falsifiable predictions. In the setting of \Cref{sec:s}, quantum theory based on \eqref{eq:mf} predicts that Wigner's experience is compatible with the measurement outcome of $\ketbra{\Phi}$ with \eqref{eq:Phi}, while collapse models rule out such coherent macroscopic superposition of Friend in accounting for Wigner's experience.

To people who object \eqref{eq:mf} on philosophical grounds for failing to give a collective account, or for being ``soliptic'', the response is that these objections are insignificant. A theory can be true or false (e.g., passing or failing the kind of empirical test discussed above) independent of whether or not it gives a collective account, and whether or not it is ``soliptic''.

\subsection{Presence vs. absence}\label{sec:pva}



Some ``realists'' aspire to understand stable set of properties of matter in and of itself, without regard to human perceptions \cite{Smolin2019EinsteinsRevolution}. A path integral like \eqref{eq:pib} does suggest a very simple picture of what Nature is like, when no experiential beings are present. Without experiential beings, there is no experience selection. All path integral configurations for matter and gravity coexist in superposition. 

The problem is that this picture, or any picture without experiential beings, is not empirically verifiable. It is not empirically verifiable because there is no experience without experiential beings. If someone gives a different picture in which the universe becomes a banana whenever experiential beings are absent, we cannot give empirical evidence to rule out the case. In \Cref{sec:d} an alternative world picture is offered, which takes as a starting point the presence of experiential beings.


As an alternative, a ``realist'' may ascertain the presence of experiential beings, and aspire for a world picture that admits a collective account of experiences in the sense of \Cref{sec:ic}. As explained there, quantum theory does not meet this hope. Whether this is good or bad should be decided by empirical tests, such as the kind mentioned in \Cref{sec:ic}, which distinguish theories that come with collective accounts and theories that do not. 

\subsection{Boundary condition and experience selection vs. state}\label{sec:bcs}


The account of Wigner's friend setting of \Cref{sec:wf} touches base with fundamental theories with specified dynamical laws encoded in the path integral action. In \Cref{sec:tm} an explicit explanation is given on how to start from formula \eqref{eq:mf} for the fundamental path integral to arrive at the formulas of \Cref{sec:wf} for the toy models. This explanation clarifies that quantum states evolving in time have no status in the fundamental path integral \eqref{eq:mf},\footnote{See, e.g., \cite{Hartle1995SpacetimeSpacetime, Sorkin1997ForksGravity, Oeckl2019APhysics} for related discussions.} which only refers to the boundary condition of the universe and experience selections. States evolving in time arise only after imposing simplifying assumptions.

What is the nature of the quantum state? Does it describe our knowledge of reality, or reality itself? Much discussion in the literature treat these as essential questions for understanding quantum theory. Yet from the perspective of the fundamental path integral they are not essential questions, because the state is a dispensable concept. 

What about the nature of the boundary condition and the experience selections? Do they describe our knowledge of reality, or reality itself? In \eqref{eq:mf}, the boundary condition is kept fixed for all experiences, while the experience selection varies for different experiences. To the extent that experience is part of ``reality'', both are used in describing ``reality''. Neither describes our knowledge \textit{per se}, although knowledge can be formed about the boundary condition and the experience selections.

\section{On quantum interpretations}\label{sec:qi}
 
What does the analysis of experience in theories of everything inform us about quantum interpretations? In this section I explain why judged in this context of theory of everything, all the interpretations listed in \Cref{tab:interpretations} face major issues.
 
\begin{table}
\centering
\begin{tabular}{|l|c|c|c|c|}
\hline
                     & Wrong & Redundant & Vague & Superfluous \\ \hline
Pilot-wave theories    &   \checkmark 
&           &       &              \\ \hline
Collapse models      &   \checkmark 
&           &       &              \\ \hline
Everettian           &   \checkmark 
&           &       &       \checkmark       \\ \hline
Decoherent histories &       &    \checkmark 
&       &       \checkmark       \\ \hline
Relational           &       &    \checkmark 
&       &    \checkmark          \\ \hline
QBism                &       &           &  \checkmark      &       \checkmark       \\ \hline
Neo-Copenhagen        &       &           &   \checkmark     &    \checkmark          \\ \hline
\end{tabular}
\caption{Dangers faced by interpretations, judged in the context of theory of everything}
\label{tab:interpretations}
\end{table}

\subsection{Wrong?}

To fulfill the first task for a theory of everything (\Cref{sec:i}), a putative theory should offer the correct laws for matter and gravity. The high danger in pilot-wave theories \cite{Teufel2009BohmianTheory} and collapse models \cite{Bassi2013ModelsTests} is that they supply the wrong laws for matter. The high danger in Everettian interpretations is that they supply the wrong laws for gravity.

\subsubsection{Pilot-wave theories and collapse models}

For laws of matter, the best theory we have is the Standard Model, which passes stringent tests from particle physics experiments. The currently available pilot-wave theories and collapse models cannot correctly reproduce the Standard Model predictions \cite{Wallace2020OnProblem, WallaceTheEvidence}. Future will tell if further developments can overcome this challenge.

\subsubsection{Everettian interpretations}

The goal of Everett's original paper \cite{Everett1957RelativeMechanics} was to present ``a reformulation of quantum theory in a form believed suitable for application to general relativity''. Back in 1957, far less is known about quantum gravity, and Everett chose to base his interpretation on the Schr{\"o}dinger equation (instead of path integral or the Wheeler-DeWitt equation, which is only available later in history). Nowadays we understand much better that unless Nature very unexpectedly singles out some time parameter in quantum gravity, a theory based on Schr{\"o}dinger equation will be wrong \cite{oriti2009approaches, kiefer2007quantum}. Therefore the Everettian view that a wave function evolving under the Schr{\"o}dinger equation is all there is \cite{2010ManyReality} is quite likely wrong for the laws of gravity.

One possible rescue is to resort to decoherent histories, which does not rely on the Schr{\"o}dinger equation \cite{Hartle1995SpacetimeSpacetime}, to formulate some alternative ``Everettian interpretation'' \cite{Wallace2012TheInterpretation} in variation from Everett's original one. However, decoherent histories also face issues, as discussed in \Cref{sec:dh}. In addition, present accounts of the Everettian interpretation in decoherent histories \cite{Wallace2012TheInterpretation} are unjustified in imposing the branching structure, due to its time-oriented nature \cite{JiaDecoherenceBranching}. Additional issues are discussed and debated in \cite{2010ManyReality}.

\subsection{Redundant?}

Decoherent histories \cite{Hartle1995SpacetimeSpacetime, Gell-Mann2013AdaptiveRealms} and Relational quantum mechanics \cite{Laudisa2021RelationalMechanics, Rovelli2022TheInterpretation} have no trouble accommodating the Standard Model or modern theories of quantum gravity (see \cite{Hartle1995SpacetimeSpacetime} for decoherent histories; Chapter 2 of \cite{Rovelli2014CovariantGravity} and Section 43.9 of \cite{Rovelli2022TheInterpretation} for Relational quantum mechanics) However, they face the danger of redundancy.

\subsubsection{Decoherent histories}\label{sec:dh}

Gell-Mann and Hartle hold that: ``The most general objective of quantum theory is the prediction of the probabilities of individual members of sets of alternative coarse-grained time histories of the closed system'' \cite{Gell-Mann2013AdaptiveRealms}. In their decoherent histories interpretation \cite{Hartle1995SpacetimeSpacetime, Gell-Mann2013AdaptiveRealms}, probabilities are assigned to sets of histories obeying the decoherence condition, which ensures the usual probability sum rules under coarse-graining. The histories do not have to refer to experiences, but could, for instance, refer to properties of matter in the early universe before any experiential being came into being. As it currently stands, the decoherent histories interpretation is unable to offer unambiguous predictions. This is due to the lack of a history selection criteria \cite{Dowker1996OnMechanics}: generically a given history can be embedded in many decoherent sets of histories, and the interpretation offers no rule for selecting among the possibilities. Without a fixed way to extend histories, probabilistic predictions cannot be made unambiguously even for the simple question of what happens next.

The analysis of the previous sections show that as far as empirical predictions are concerned, probability assignment can be restricted to individual experiences instead of histories. There is still the open question to determine the set of possible experiences under a given condition (\Cref{sec:gfe}). It is reasonable to hope that a theory of experience can address this question, because if there is going to be a scientific theory of experience at all, as a basic requirement it should tell us what experiences are possible under a given condition.

In contrast, a theory of experience will not reduce the ambiguities associated with decoherent histories. Firstly, as a defining feature, the decoherent histories interpretation encompasses both experiential and non-experiential histories. A theory of experience will not help to reduce the ambiguities associated with non-experiential histories. Secondly, even if one restricts attentions to experiential histories, they will generically refer to multiple experiences. As noted in \Cref{sec:woe}, ``joint experience is not experience''. It is unclear how a theory of experience can fix ambiguities associated with histories of such non-experiences. 

In particular, the ambiguities raised in \Cref{sec:eqp} appear here as well for experiential histories. Should one assign probability to the histories of multiple experiences (which could be for one being or for multiple beings)? Which set of histories should one use, if one just wants to predict Alice's next experience? One possible way out is to follow the logic of \Cref{sec:eqp} and assign probabilities only to individual experiences, but not histories. This would suffice to connect theory with experience to meet Task 3 of \Cref{sec:i}. In this case, assigning probabilities to histories would be fundamentally redundant. This is not to say that history considerations should be forbidden. Instead, they can still be of practical utility, for as noted by Hartle \cite{Hartle1998QuantumHistory}: ``In principle the same prediction could be made from the present data themselves [...] However, it is evidently much easier to start from the event in the past. The reason is that present data contain much information that is irrelevant for this particular future prediction.''

\subsubsection{Relational Quantum Mechanics}

Relational quantum mechanics (RQM) \cite{Laudisa2021RelationalMechanics, Rovelli2022TheInterpretation} considers ``facts'' (physical variable taking definite value), which take place whenever two systems interact. Here the systems do not have to be experiential beings, but can be any physical system. Facts are relative to the systems that interact, and are labelled by the interacting systems.

We saw from \Cref{ovs} that with the minimal prescription, quantum theory allows objective empirical predictions: Everyone uses a common formula to predict anyone's experiences. An experience is labelled by the experience itself ($e_i,c$ of \eqref{eq:mf}), and nothing else. In particular, there is no need to refer to systems and their interactions. From this view, RQM is redundant in introducing relative labels where no relative label is needed.

One might argue that there is a gain in the RQM move, because by assigning definite values to variables whenever systems interact, it treats empirical facts and non-empirical facts on equal footing, so that it avoids singling out empirical facts as special. However, in RQM this move only undermined the interpretation itself. For theories of the kind of \Cref{sec:i}, RQM holds that ``variables actualize at three-dimensional boundaries, with respect to (arbitrary) spacetime partitions'' (Section 43.9 of \cite{Rovelli2022TheInterpretation}). Here an arbitrary spacetime partition single out two systems, and the interaction at the three-dimensional boundary is supposed to trigger variable actualization at the boundary.
Indeed, this does not draw a distinction between empirical facts and non-empirical facts, but it begs the question of whether there are empirical facts at all. Why should one believe that experiences correspond to variables taking definite values on 3D surfaces around 4D regions? Why do experiences not correspond to variables taking definite values, for instance, within 4D regions themselves? Do experiences correspond to variables taking definite values throughout the 3D surfaces? Why not on part of the surfaces? How many 3D surfaces should one consider, for one experience of Alice? ... Without addressing these concerns, RQM's move of treating empirical and non-empirical facts on equal footing obscures how variable actualization relates to experience, and obstructs empirical predictions from the interpretation. 

Another side effect of the move is that RQM has to face the preferred basis problem \cite{Mucino2022AssessingMechanics, BruknerQubitsTheorem}: An interaction does not in general single out a basis for physical variables to take definite values, so with respect to which basis do variables take definite values? Curiously, partially in response to this problem, Adlam and Rovelli now considers a new version of RQM which contains special suppositions for conscious observers \cite{AdlamInformationMechanics}: ``Therefore in this version of RQM it is now feasible to suppose that the perspective of a conscious observer simply emerges from the collection of the perspectives of all the particles in their brain - roughly speaking, a variable $V$ of a system $S$ will have a definite value $v$ relative to me if variable $V$ has the definite value $v$ relative to most of the particles in my brain (or perhaps just in some particularly relevant section of my brain - we would have to turn to neuroscience to determine how much of the brain should be included).'' On the preferred basis problem, Adlam and Rovelli remark that: ``RQM need only show that in the limit as one of the systems involved becomes macroscopic, then there is a unique choice of variable which takes definite values in the interaction, in order that macroscopic conscious beings like ourselves can have definite experiences.'' Decoherence is supposed to pick out the basis, even if ``the decoherence process is not perfectly well-defined - there is no exact line between ‘decohered’ and ‘non-decohered’'', because ``consciousness also does not seem to be perfectly well-defined: to our best current understanding it appears to be some kind of emergent high-level feature of reality, so we are certainly entitled to suppose that consciousness can emerge only when enough decoherence has occurred to single out a well-defined preferred basis.'' In this new version of RQM to address the preferred basis problem, empirical and non-empirical facts are clearly not on equal footing, so there is no longer any gain in this aspect, in comparison to the minimal prescription of \Cref{ovs}. One wonders if there is any gain at all in the RQM move to compensate the redundancy of introducing relative labels discussed above.

    

\subsection{Vague?}

QBism \cite{Fuchs2016QBism:Handbook, StaceyIdeasQBism} and neo-Copenhagen \cite{Brukner2017OnProblem} interpretations do not assign probabilities to non-empirical histories or facts, so avoids the issue of redundancy. However, they are formulated in vague terms that hinder a direct application to fundamental theories like those in \Cref{sec:i}.

\subsubsection{QBism}

QBism \cite{Fuchs2016QBism:Handbook, Fuchs2017NotwithstandingQBism, StaceyIdeasQBism} holds that quantum measurement outcomes are just personal experiences for the agents, and quantum probabilities are subjective personal degrees of belief for agents. 

QBism has been criticized for being vague and ambiguous (see \cite{Zwirn2022IsMechanics} and references therein). For example, the central concept of agent is loosely characterized as ``entities that can take actions freely on parts of the world external to themselves, so that the consequences of their actions matter for them'' \cite{DeBrota2020RespectingFriend}. This engenders some ambiguities. What counts as ``take actions freely''? Does a cell count as an agent? A robot? Alice's brain? Alice's brain in conjunction with Bob's legs? On what basis? One possible way out is to identify individual agents with individual experiential beings. Yet given that joint experience is not experience (\Cref{sec:woe}), this would contradict some previous QBism understandings, e.g., a collection of scientists can act as a single agent (p.1872 of \cite{DeBrota2020RespectingFriend}). Ambiguities like this associated with basic notions leave one in wonder what QBism really is saying.

In the context of this paper, another critical issue is whether QBists' focus on the quantum state $\psi$ is misguided. In the understanding of \Cref{ovs}, although experiences themselves are subjective, empirical predictions based on \eqref{eq:mf} are objective, in the sense that everyone should use the same formula to predict anyone's experiences. This point is made clear by analyzing the path integral formula \eqref{eq:mf}, which refers directly to the fundamental gravity and matter variables. In contrast, QBism focuses on quantum states $\psi$, which have no fundamental status in the sense explained in \Cref{sec:bcs}. If one is free to pick one's favourite $\psi$ to bet on future experiences, certainly some choices will help one gain money, while some will not. It is not false to say that $\psi$ and probabilities derived from it are subjective, but this observation is quite unimportant, if there is formula which always gives the correct empirical predictions to win the bets. \Cref{eq:mf} is supposed to be such a formula, which also makes objective predictions, in apparent tension with QBism. 

\subsubsection{Neo-Copenhagen interpretation}

Brukner's neo-Copenhagen interpretation \cite{Brukner2017OnProblem} ascertains the object-subject cut of the traditional Copenhagen interpretation by taking measurement instruments to ``lie outside the domain of the theory'', and holds that ``the quantum state is a representation of knowledge of a (real or hypothetical) observer relative to her experimental capabilities [...] The available experimental precision will in every particular arrangement determine to which objects the observer can meaningfully assign quantum states.''

It is not clear how this interpretation applies to theories like those in \Cref{sec:i}. What constitutes an ``observer'' in terms of the fundamental matter and gravity variables? What about an  ``experiment''? Is there a distinction between experiment and everyday experiences? Does the interpretation accept \eqref{eq:mf} for empirical prediction? If not, what formula does it give? In the brief reference to quantum cosmology in \cite{Brukner2017OnProblem}, it is mentioned that the observer ``is always considered to be external to the universe'', since ``the `wave function of the universe' that would include the observer is a problematic concept, as it negates the necessity of the object–subject cut''. Does \eqref{eq:mf} with its empirical selection count as treating the observer (or experiential being) external or internal to the universe? It is difficult to infer a definitive answer from \cite{Brukner2017OnProblem}.

The key notion of ``fact'' especially deserves a clarification in the neo-Copenhagen interpretation. In the discussion of Wigner's friend settings \cite{Brukner2018AFacts}, ``facts'' which are assigned quantum probabilities can refer to both ``measurement records'' and ``immediate experiences of observers'', and in some statements, ``facts'' is used interchangeably with ``measurement records'' and ``experiences'' \cite{Brukner2017OnProblem}: ``I will show that any attempt to assume that the measurement records (or “facts” or experiences) that coexist for both Wigner and his friend will run into the problems of the hidden variable program...'' However, there is a big difference between record and experience. While joint records is still a record, joint experiences is no longer an experience (\Cref{sec:eqp}). If probabilities are assigned also to measurement records even when no experience refers to them, one is left in wonder what count as records for theories like those in \Cref{sec:i}.

Like QBism, the neo-Copenhagen interpretation focuses on quantum states $\psi$, which have no fundamental status in the sense explained in \Cref{sec:bcs}. This is another gap to be filled before one could apply the interpretation to the fundamental theories with specified dynamical laws, if it can be applied to fundamental theories with specified dynamical laws at all.

\subsection{Superfluous?}


The above dangers of being wrong, redundant, or vague are possibly avoided with additional inputs. Yet a much more important issue needs to be tackled first. Are these interpretations worth developing further at all? If the goal is to find a satisfactory physical theory of everything, what difference do the above interpretations make toward this goal? 

Alternative theories such as pilot-wave theories and collapse models make different empirical predictions from quantum theory. They also allow for a collective account of experiences in the sense of \Cref{sec:ic}, such that if they succeed, we gain a totally different worldview from quantum theory. These differences they make make them worth developing.

The situation is far less clearer for the other interpretations discussed above. Consider the three tasks given in \Cref{sec:i} for a theory of everything. The interpretations do not tell us the correct dynamical laws for matter and gravity, nor the correct laws for the boundary condition of the universe. The analysis of \Cref{sec:eqp} gives formula \eqref{eq:mf} to relate theory with experience, without subscribing to any of these interpretations. The outstanding problem is to find a theory of experience which can address the open questions listed in \Cref{sec:gfe}. In their current form, these interpretations seem to make no contribution to this task, either. Without contributing anything to the physical theory of everything, these interpretations face the real danger of being superfluous.

\section{Discussion}\label{sec:d}

Theories of physics are often derived and tested in controlled laboratory settings. This does not mean that the theories only apply to controlled laboratory events. We have learned that physical laws of gravity govern not only Galilei's wood and iron balls, but also birds in free flight. We have learned that physical laws of electromagnetism govern not only Faraday's coils of wire, but also lightnings in thunderstorms.

In this paper, I considered the case that quantum theory applies not only to microscopic phenomena in controlled laboratory settings, but also to everyday experiences. It is assumed that experiences take place probabilistically, i.e., an experience-enabling physical condition $c$ makes possible a set of experiences $\{e_i\}_i$ to take place with certain probabilities. In path integral theories of everything, the probabilities are given by formula \eqref{eq:mf}, reproduced here as
\begin{align}
    p(e_i|c)=N \int_{e_i,c} Dq'  \int_{e_i,c} Dq ~ 
e^{\frac{i}{\hbar}\left(S[q] -S[q']\right)}\rho(q_b, q_b').\nonumber
\end{align}
The path integral is doubled to incorporate possibly mixed boundary conditions $\rho$. The sums include all gravity and matter configurations constrained by the condition $c$ and the experience $e_i$ it enables.\footnote{The constraint may take the form of a characteristic function, so that gravity and matter configurations compatible with $c$ and $e_i$ are included, while other configurations are excluded. The constraint may also take the form of a function which assigns complex weights to the gravity and matter configurations according to $c$ and $e_i$, e.g., in implementing ``unsharp measurements'' induced by the experience. As discussed in \Cref{sec:gfe}, the precise form of the constraint hinges on some yet unavailable scientific knowledge about the relationship between physics and experiences.}

The world picture that emerges from formula \eqref{eq:mf} can be counter-intuitive to some. In accounting for one experience $e_i$ with condition $c$, we must take it that everywhere else all matter and gravity path integral configurations coexist in superposition. For instance, in the Wigner's friend setting, all other beings must be put in superposition when accounting for one being's experience. This world picture is forced upon us, since by the analysis of \Cref{sec:woe}, the other prescriptions which select path integral configurations differently are not justifiable.

How are different experiences related, if each use of \eqref{eq:mf} only refers to one experience? For the sequential experiences of one being, suppose that previously the experience $e_i$ took place under condition $c$. This gives rise to another condition $c'$ that enables another set of experiences $\{e'_j\}_j$, whose probabilities are given by \eqref{eq:mf} for $p(e'_j|c')$. The pair $(e_i',c')$ in turn gives rise to another condition with another set of experiences, etc. Considerations of other beings' experiences are motivated by the content of the experiences of the first being. If in the experience of one being, it perceives the presence of other experiential beings, then it can guess at the conditions $d$ and experiences $\{f_i\}_i$ of the other beings and draw probabilistic predictions for the other beings' experiences using \eqref{eq:mf}. 

Strictly speaking, no consideration of other beings is needed to account for the experiences of one being. This is implied by \eqref{eq:mf} which holds that all experiences are accounted for by selecting just for that experience. However, in practice we often apply multiple selections in a third-person view (e.g., we apply measurement projections to multiple parties for a Bell experiment) and this often yields valid empirical predictions. How come? The answer is the same as why we often apply non-relativistic analysis which often yield valid empirical predictions. Often a non-relativistic analysis already yields results that meet the desired accuracy. A more consuming relativistic analysis would be an overkill. Likewise, often a third-person view quantum analysis already yields results that meet the desired accuracy. In this case a more consuming first-person analysis based on \eqref{eq:mf} could be saved.

On the other hand, Relativity does not always agree with non-relativistic physics, and it is important to figure out when this happens. Likewise, quantum theory according to \eqref{eq:mf} does not always agree with naive prescriptions, and it is important to figure out when this happens as well. Wigner's friend setting of \Cref{sec:wf} is one prominent example. There should be more examples waiting to be explored. For instance, in cosmology, taking \eqref{eq:mf} as the basic formula for empirical predictions will provide a basis to examine how inhomogeneities and anisotropies arise \cite{Castagnino2017InterpretationsCosmology}, as well as to resolve the Boltzmann brain problem \cite{Carroll2017WhyBad}. Specifically, since an experience is characterized completely by the local selection $e_i,c$ of \eqref{eq:mf}, the assumption of distinguishability between ordinary and Boltzmann brain experiences based on gravity configurations outside the brains cannot be retained. In another realm, for measurements in QFT, \eqref{eq:mf} enforces a first-person description, which can clarify the issue of superluminal signalling faced by naive measurements prescriptions in a third-person description \cite{Sorkin1993ImpossibleFields, Borsten2021ImpossibleRevisited, Jubb2022CausalTheory}. Specifically, we must wait for the other agents to send back signals to the agent whose experience is under consideration to distinguish superluminal and non-superluminal cases. 

In this paper, I considered \eqref{eq:mf} for arbitrary selections $c$ and $e_i$. As noted in \Cref{sec:gfe} there are open questions to address if one wants to be more explicit about what  $e_i$ and $c$ correspond to actual experiences and conditions. What physical conditions $c$ enable experience? What determines the set of possible experiences $\{e_i\}_i$ given $c$? What determines the next condition $c'$ given the previous experience $e_i$ and condition $c$? Are there lawlike features yet to be uncovered about experience? What is the status of ``Free Will'' in this context? ... Addressing these questions will most require interdisciplinary efforts that push further the boundaries of physics \cite{Maturana1980AutopoiesisCognition, Capra2012TheVision, Auletta2011CognitiveMinds, Bialek2012Biophysics:Principles, Kent2000NightPhysicist, JiaModesWorld}. 

\section*{Acknowledgement}

I thank Chris Fuchs for some kind comments (with Emoji) about QBism on an earlier draft. I am very grateful to Lucien Hardy and Achim Kempf for long-term encouragement and support. Research at Perimeter Institute is supported in part by the Government of Canada through the Department of Innovation, Science and Economic Development Canada and by the Province of Ontario through the Ministry of Economic Development, Job Creation and Trade. 


\printbibliography 

\end{document}